\documentclass[aps,prl,twocolumn,superscriptaddress,showpacs]{revtex4}

\usepackage{graphicx}

\begin{document}

% Use the \preprint command to place your local institutional report
% number in the upper righthand corner of the title page in preprint mode.
% Multiple \preprint commands are allowed.
% Use the 'preprintnumbers' class option to override journal defaults
% to display numbers if necessary
%\preprint{}

%Title of paper
\title{
  A new screening function for Coulomb renormalization
}

% repeat the \author .. \affiliation  etc. as needed
% \email, \thanks, \homepage, \altaffiliation all apply to the current
% author. Explanatory text should go in the []'s, actual e-mail
% address or url should go in the {}'s for \email and \homepage.
% Please use the appropriate macro foreach each type of information

% \affiliation command applies to all authors since the last
% \affiliation command. The \affiliation command should follow the
% other information
% \affiliation can be followed by \email, \homepage, \thanks as well.
\author{
M. Yamaguchi
}
\email{yamagu@rcnp.osaka-u.ac.jp}
%\homepage[]{Your web page}
%\thanks{}
%\altaffiliation{}
\affiliation{
Research Center for Nuclear Physics, Osaka University,
Ibaraki 567-0047, Japan
}

\author{
H. Kamada
}
\email{kamada@mns.kyutech.ac.jp}
\affiliation{
Department  of Physics, Faculty of Engineering, Kyushu Institute of Technology,
 Kitakyushu 804-8550, Japan
}

\author{
Y. Koike
}
\email{koike@i.hosei.ac.jp}
\affiliation{
Science Research Center, Hosei University, 2-17-1, Fujimi, Chiyoda-ku,
Tokyo 102, Japan
}

\date{\today}

\begin{abstract}
We introduce a new screening function which is useful for
the few-body Coulomb scattering problem in ``screening
and renormalization'' scheme.
The new renormalization phase factor of the 
 screening function is analytically shown.
The Yukawa type of the screening potential has been used 
in several decades, we modify it to make more useful.
As a concrete example, we compare the proton-proton scattering phase shifts
calculated from these potentials.
The numerical results document that high precision
 calculations of the renormalization are performed by the 
new screening function.
\end{abstract}

% insert suggested PACS numbers in braces on next line
\pacs{24.10.-i,21.45.+v,25.40.Cm,02.60.-x}
% insert suggested keywords - APS authors don't need to do this
%\keywords{}

%\maketitle must follow title, authors, abstract, \pacs, and \keywords
\maketitle

%\narrowtext
When one  considers the system of few  charged 
 particles, one faces, as is well known,
 serious difficulties in the calculation of
 scattering processes\cite{alt-text}. This is caused by the long range 
nature of  the Coulomb potential in
 coordinate space, or, equivalently, its singularity
 in momentum space.

In order to overcome  these  difficulties, we would like to mention 
two approaches.
One approach is a modified time-dependent scattering theory\cite{dollard64}
 and another approach is a ``screening and renormalization'' method.
Many investigations have been done in  few-body systems by
 the latter approach.
For example, Alt {\it et. al.} \cite{alt2002}
have calculated  three-body scattering
 with  charged particles by the screening
 and renormalization scheme.
Their calculations worked out successfully.
However, the screening radius $R$  used in their calculation is about 600
 fm. 
At such a large R-value the screened Coulomb potential is no longer smooth
requiring a careful treatment and increased computer resource in memory 
and elapsed time. Therefore, it would be desirable to work with a smaller 
R-value. The purpose of this letter is to investigate a new screening 
function different from the one used in  Alt {\it et. al.} which leads to
precise results even at small R-values.    

Before we discuss the  new screening function,
 we would like to point out the significance of the renormalization.
If one considers a bound state numerical calculations can be performed 
by the screening method with an appropriately large R-value and the result
will be independent of the choice of R. However, for a scattering state
the situation is completely different. The limit of solutions achieved with 
screened Coulomb potentials for increasing R-values will not agree with the 
solutions for a pure Coulomb potential. The reason lies in the wrong 
asymptotic boundary conditions going with a screened Coulomb potential.
A renomalization method, like the one introduced by Taylor \cite{taylor74}
is necessary. 
 
Alt {\it et. al.} used the Yukawa type screened Coulomb potential
 with the screening radius $R$:

 \begin{eqnarray}
   V^{R} (r) = e^2 \frac{\exp[-(r/R)]}{r}.
   \label{eq:1}
 \end{eqnarray}
Here in this paper we have proton-proton scattering in mind,
 therefore, $e$ represents the charge of the proton.
Now we introduce the new screening functions:
 \begin{eqnarray}
   V^{R}_{n} (r) = e^2 \frac{\exp[-(r/R)^n]}{r}.
   \label{eq:2}
 \end{eqnarray}
Note that eq.(\ref{eq:2}) reduces to the Yukawa type potential
 for $n=1$ and is also going to
a  sharply truncated potential for $n \rightarrow \infty$,
 \begin{eqnarray}
   \lim_{n \rightarrow \infty} V^{R}_{n} (r) =
   \frac{e^2}{r} \cdot \theta(R-r),
   \label{eq:3}
 \end{eqnarray}
 where $\theta(r)$ is the $\theta$-function.
The pure and the screened Coulomb potentials for $n$ = 1 to 5 at
 $R$ = 50 fm are shown 
in Fig. 1. This figure reveals that with large n there develops a sharper 
cut-off and the pure Coulomb potential is better represented in the inner 
region.   
\begin{figure}[hbt]
\begin{center}
 \includegraphics{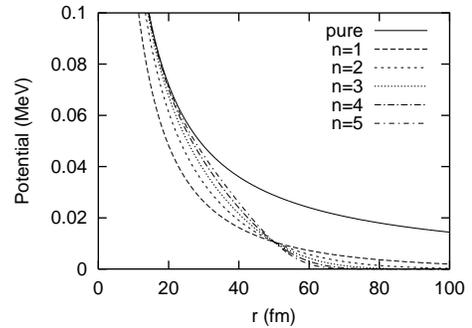}
\end{center}
\caption{Pure and screened Coulomb potentials for $n = 1-5$.
 }
\label{fig:1}
\end{figure}

Next, we evaluate the phase shift for pp scattering in the state $^1S_0$.
As a typical NN force we take the Reid Soft Core
potential\cite{reid68}
 in addition to the
Coulomb potential. There exists the following relation\cite{taylor74}
among various phase shifts, which can be gained by regarding the asymptotic 
behavior of the wave function:

 \begin{eqnarray}
   \delta^C = \delta^R - \sigma_0 + \phi_R + 0 ( 1 / R )
   \label{eq:4}
 \end{eqnarray}
Here  $\delta^C$ is the phase shift to the strong and pure Coulomb potential,
  $\delta^R$ to the strong and screened Coulomb potential, 
 $\sigma_0$ the standard Coulomb phase shift obtained from
  $\arg \Gamma (1 + i \eta )$ and  $\phi_R$ the renormalization
   phase.
 Further, 
 $\eta = e^2 m / 2p$ is Sommerfeld parameter with $m$  the nucleon mass and   
$p$  the relative momentum. According to Taylor 
the renormalization phase $\phi_R (p)$ is given as 
 \begin{eqnarray}
   \phi_R (p) &\equiv& -\eta \int^{\infty}_{(2p)^{-1}}
                        \frac{\exp[-(r/R)^n]}{r} dr   \nonumber \\
   &=& -\eta \int^{r}_{0} 
       \frac{\exp[-(r/R)^n] - 1}{r} dr -\eta \ln (2pr) \nonumber \\
   && \quad   + 0 ( 1 / R^n )                          \nonumber \\
   &=& -\eta \left[ \ln (2pR) - \frac{\gamma}{n} \right]
       + 0 ( 1 / R^n ),
 \label{eq:5}
 \end{eqnarray}
 where $\gamma$ is the Euler number $(0.5772\cdots)$.
The last line in Eq. (5) results using the substitution 
 $ \left( r/R \right)^n = s/R $. For n=1 this is the result used by Alt  {\it
 et.al.}. For $ n \to \infty$ one obtains  $\phi_R = -\eta \ln ( 2 p R )$, 
which is the expression related to a sharp cut-off.
Note that the definition of $\phi_R (p)$ suffer from the uncertainty
 of $0 ( 1 / R ) $ in eq. (\ref{eq:4})

\begin{figure}[hbt]
\begin{center}
 \includegraphics{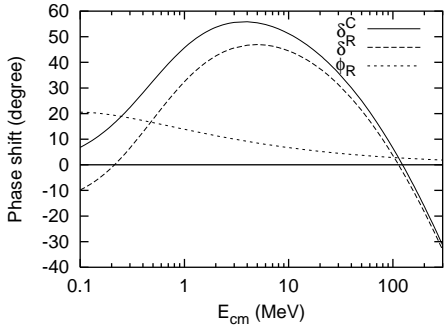}
\end{center}
\caption{The comparison of the phase shifts $\delta^C$, $\delta^R$
 and $\phi_R$ for $R=50$ fm.
}
\label{fig:2}
\end{figure}

The phase shifts  $\delta^C$, $\delta^R$ and $\phi_R$ are shown
 in Fig.\ref{fig:2}.
In the evaluation of the phase shifts one has to study the dependencies of the 
c.m. energy $E_{cm}$, the screening radius R and the power n in our new 
screening function. As a measure for the quality of the new screening function
we introduce 
 \begin{eqnarray}
 | \Delta \delta | \equiv \ | \ \delta^C - \delta^R - \eta[\gamma / n
 - \ln (2 p R)] \ | \ 
  \label{eq:6}
 \end{eqnarray}

\begin{figure}[hbt]

\begin{center}
 \includegraphics{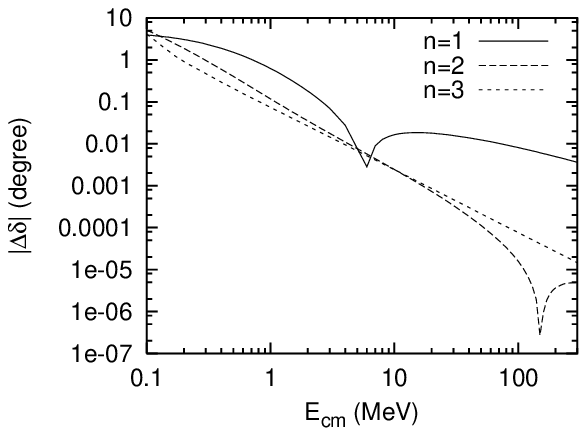}
\end{center}
\caption{ $ | \Delta \delta | $  for $R = 50$ fm against 
$E_{cm}$ for various n-values.
 }
\label{fig:3}

\begin{center}
 \includegraphics{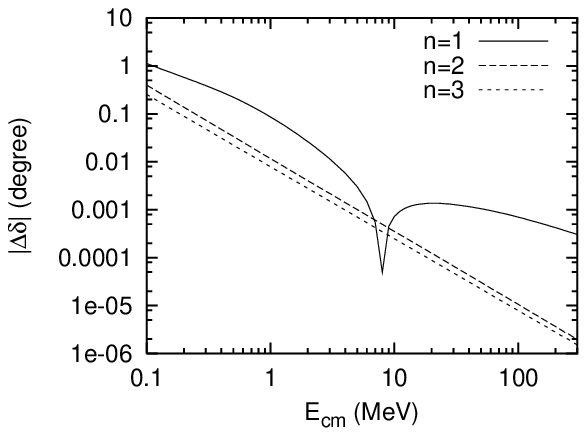}
\end{center}
\caption{The same as in Fig. 3 for $R = 500$ fm.
}
\label{fig:4}
\end{figure}

\begin{figure}[hbt]

\begin{center}
 \includegraphics{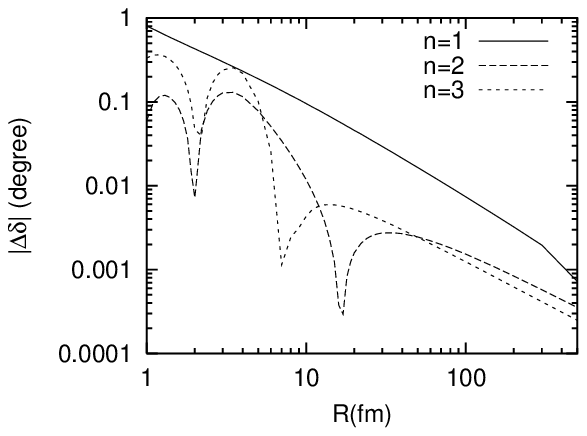}
\end{center}
\caption{ $| \Delta \delta |$ for $E_{cm}=10$ MeV against R 
for various n.
}
\label{fig:5}

\begin{center}
 \includegraphics{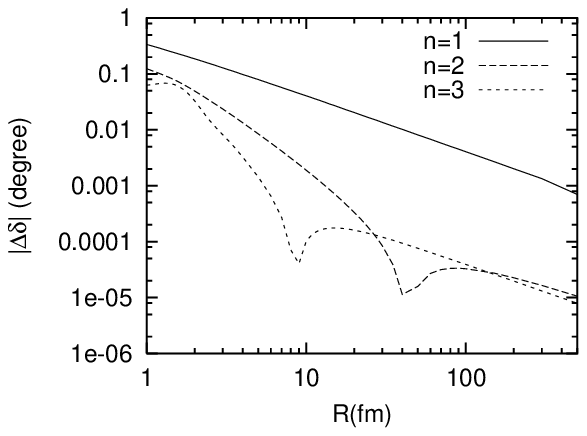}
\end{center}
\caption{The same as in Fig. 5 for  $E_{cm}=100$ MeV.
}
\label{fig:6}
\end{figure}

\begin{figure}[hbt]

\begin{center}
 \includegraphics{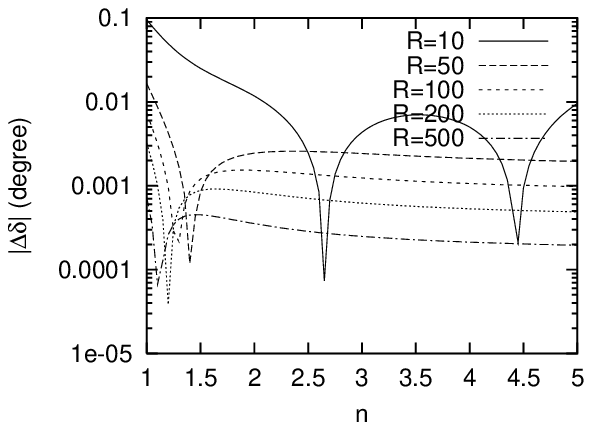}
\end{center}
\caption{ $|\Delta \delta |$ for $E_{cm} = 10$ MeV against n for various
 R-values.
}
\label{fig:7}

\begin{center}
 \includegraphics{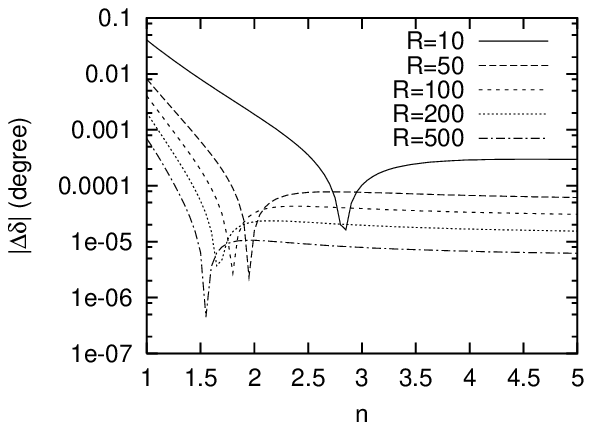}
\end{center}
\caption{The same as in Fig. 7 for $E_{cm} = 100$ MeV.
}
\label{fig:8}
\end{figure}

This quantity is plotted in Figs.\ref{fig:3}-\ref{fig:8}.  
In Fig.\ref{fig:3} and Fig.\ref{fig:4} we show 
the $ | \Delta \delta |$-dependence 
 of $E_{cm}$ for $R$ = $50$ fm and $500$ fm, respectively.
$ |\Delta \delta  |$ decreases strongly with  $E_{cm}$. 
These figures clearly show that the calculations 
for  $n \geq 2$ are more precise than for n=1. 
In view of Fig. 1 we have to conjecture that this is due to the better 
approach of the pure Coulomb potential by  the screened Coulomb potential 
in the inner region if n is larger than 1. 

In Figs. 5 - 6, we illustrate the R-dependence of  $ | \Delta \delta |$
for two fixed $E_{cm}$-values of 10 and 100 MeV. 
Trivially, $ | \Delta \delta |  $ decreases with increasing R but again 
the error is significantly smaller if n is larger than 1 in comparison to 
n=1. 

Finally Figs. 7 and 8 give the n-dependence of  $ | \Delta \delta |  $ 
for fixed $E_{cm}$ and various R's at $E_{cm}$= 10 and 100 MeV, respectively.
It is seen that R = 50 fm and n=2 is a good enough choice to perform the 
calculations.

In summary, we generalized the Yukawa type screening potential adopted 
by Alt et al.  \cite{alt2002} to the form given in Eq. (2). 
Already for n=2 the screened Coulomb potential is closer to the pure 
Coulomb potential in the inner region. The new expression for the 
renormalization phase $\phi _R$ is just a bridge between the Yukawa type 
and sharp cut-off potential. The numerical results document that high 
precision calculations can be performed by the new screening function for
$n \geq  2 $ and a screening radius as small as R=50 fm.

%\section*{Acknowledgment}

\begin{acknowledgments}
The authors would like to thank Prof. W. Gl\"ockle 
for critically reading the manuscript.
One of Authors (M.Y.) acknowledges the supports of the Theory Group of 
Research center for Nuclear Physics in Osaka University.
\end{acknowledgments}

\end{document}